\documentclass[11pt]{article}

\usepackage{soul}

\usepackage[bottom=1.0in]{geometry}

\usepackage{float}
\usepackage[dvipsnames]{xcolor}
\usepackage[colorlinks=true, linkcolor=BrickRed, urlcolor=BrickRed, citecolor=BrickRed, anchorcolor=blue]{hyperref}
\usepackage{wrapfig}
\usepackage{comment}
\usepackage{tikz}
\usetikzlibrary{positioning,arrows.meta}
\usepackage{textgreek}
\usepackage{url}

\usepackage{enumitem, comment, xifthen}
\usepackage{amsmath,amssymb,amsthm, amsfonts, bbm}
\usepackage[T1]{fontenc}
\usepackage{makecell}
\usepackage{booktabs}
\usepackage{natbib}

\usepackage[font=small]{caption}

\usepackage{booktabs}
\usepackage{hhline}
\usepackage{array,multirow}
\usepackage{lipsum}
\usepackage{siunitx,etoolbox}

\usepackage{graphicx}

\usepackage{tikz}
\usepackage{lipsum}

\let\origmaketitle\maketitle
\def\maketitle{
  \begingroup
  \def\uppercasenonmath##1{} 
  \let\MakeUppercase\relax 
  \origmaketitle
  \endgroup
}

\newcommand{\upstairs}[1]{\textsuperscript{#1}}
\newcommand{\affilone}{\dag}

\newcommand{\affilthree}{$\diamond$}

\usepackage{graphicx}
\usepackage{cleveref}
\usepackage{subcaption}

\def\beq{\begin{equation}} 
\def\eeq{\end{equation}}
\def\beqn{\begin{eqnarray*}}
\def\eeqn{\end{eqnarray*}}
\def\Bal{\begin{align}}
\def\Eal{\end{align}}
\def\Bitem{\begin{itemize}\setlength{\itemsep}{.2in}}
\def\bitem{\begin{itemize}\setlength{\itemsep}{.05in}}
\def\eitem{\end{itemize}}
\def\blatin{\begin{enumerate}\setlength{\itemsep}{.05in}\renewcommand{\labelenumi}{\roman{enumi}.}}
\def\elatin{\end{enumerate}}
\def\Benum{\begin{enumerate}\setlength{\itemsep}{.2in}}
\def\benum{\begin{enumerate}\setlength{\itemsep}{.05in}}
\def\eenum{\end{enumerate}}
\def\bmult{\begin{multline*}}
\def\emult{\end{multline*}}
\def\bcenter{\begin{center}}
\def\ecenter{\end{center}}
\def\bframe{\begin{frame}}
\def\eframe{\end{frame}}

\def\cL{\mathcal{L}}

\def\cN{\mathcal{N}}

\def\cQ{\mathcal{Q}}

\def\cX{\mathcal{X}}

\def\bbE{\mathbb{E}}

\def\bbN{\mathbb{N}}

\def\bbP{\mathbb{P}}

\def\ind{\mathbbm{1}}

\newcommand{\Var}{\operatorname{Var}}
\newcommand{\Cov}{\operatorname{Cov}}

\def\1{\mathbbm{1}}

\def\({\left(}
\def\){\right)}

\newcommand*\diff{\mathop{}\!\mathrm{d}}

\usepackage[textsize=tiny]{todonotes}
\setuptodonotes{inline}

\usepackage{MnSymbol}

\newtheorem{definition}{Definition}
\newtheorem{thm}{Theorem}
\newtheorem{as}{Assumption}

\theoremstyle{remark}
\newtheorem*{ex}{Example}
\newtheorem*{rem}{Remark}

\definecolor{revisered}{RGB}{210, 43, 43}
\newcommand{\revised}[1]{{\color{black}#1}}

\begin{document}



  \title{Causal Meta-Analysis: Rethinking the Foundations of Evidence-Based Medicine
   \vspace{-1em}}
  \maketitle

  \thispagestyle{empty}
  
  \begin{tabular}{cc}
    Clément Berenfeld \upstairs{\affilone}, Ahmed Boughdiri \upstairs{\affilone}, Bénédicte Colnet,  Wouter van Amsterdam \upstairs{\affilthree}, \\
     Aurélien Bellet \upstairs{\affilone},
     Rémi Khellaf \upstairs{\affilone},
     Erwan Scornet \upstairs{*},
    Julie Josse \upstairs{\affilone}
   \\[0.25ex]
   {\small \upstairs{\affilone} PreMeDICaL, Inria,
  Univ Montpellier, France} \\
   {\small \upstairs{\affilthree} University Medical Center Utrecht, Netherlands} \\
   {\small \upstairs{*} Sorbonne Université and Université Paris Cité, CNRS}
  \end{tabular}
  
 
\begin{abstract}

Meta-analysis, by synthesizing effect estimates from multiple studies conducted in diverse settings, stands at the top of the evidence hierarchy in clinical research. Yet, conventional approaches based on fixed- or random-effects models lack a causal framework, which may limit their interpretability and utility for public policy.
Incorporating causal inference reframes meta-analysis as the estimation of well-defined causal effects on clearly specified populations, enabling a principled approach to handling study heterogeneity.
We show that classical meta-analysis estimators have a clear causal interpretation when effects are measured as risk differences. However, this breaks down for nonlinear measures like the risk ratio and odds ratio. To address this, we introduce novel causal aggregation formulas that remain compatible with standard meta-analysis practices and do not require access to individual-level data.
To evaluate real-world impact, we apply both classical and causal meta-analysis methods to 500 published meta-analyses. While the conclusions often align, we unveil cases where conventional methods may suggest a treatment is beneficial when, under a causal lens, it is in fact harmful.

\end{abstract}

\vspace*{0.15in}
\hspace{10pt}

\section{Meta-analysis: the pinnacle of evidence-based medicine}
\label{sec:intro}
In contemporary evidence-based medicine, randomized controlled trials (RCTs) are widely regarded as the gold standard for estimating treatment effects, due to their unique ability to isolate treatment effects from confounding factors (see for e.g. French's health authority assessment guidelines in \cite{SanteGouvDoctrine}).
By assigning treatment randomly, much like a coin toss, RCTs ensure that, asymptotically, patient characteristics are balanced across treatment and control groups \citep{imbens2015causal}. 
For example, the average age of treated individuals is equal to that of controls, 
ensuring that any observed differences in outcomes can be attributed to the treatment itself, rather than  extraneous variables.

However, despite their internal validity, RCTs face well-documented limitations, see e.g. \citet{deaton2018reflections} for a discussion. Their high cost, lengthy timelines, operational complexity, and strict inclusion criteria often result in small, selective samples that exclude substantial portions of real-world populations, such as individuals with comorbidities, pregnant women, or other vulnerable groups. As a result, their findings frequently lack generalizability \citep{rothwell2005external} and may be of limited relevance to clinical practice and policy-making, where patient populations are markedly more diverse.

To help overcome these limitations (particularly, the issue of limited statistical power), regulatory agencies frequently rely on meta-analyses for high-stakes decisions on drug approvals and reimbursement based on efficacy \citep{nice}. 
 By synthesizing estimated effects from multiple studies conducted in diverse settings \citep{hunter2004methods, borenstein2021introduction}, meta-analysis represents the pinnacle of evidence in clinical research \citep{blunt2015hierarchies}.

 To further improve the representativeness of clinical trials, a growing body of work has focused on generalization methods. These approaches aim to predict treatment effects in external target populations by combining individual-level data from a single RCT with external data representative of the target population \citep{pearl2011transportability, stuart2011use, dahabreh2020extending, colnet2024causal, degtiar2023review}. \revised{The external data may vary in type and can originate from both RCTs or observational studies.} These methods account for treatment effect heterogeneity, i.e., the fact that treatment efficacy may vary systematically across patient characteristics, by adjusting for covariate differences between trial and target populations. In doing so, they extend the inferential reach of RCTs, offering a principled framework for estimating treatment effects beyond the original study sample.

Generalization methods are increasingly recognized not just as a complement to traditional RCTs, but as a powerful evolution of meta-analysis approaches themselves \citep{dahabreh2023efficient, rott2024causally, hong2025estimating}, offering one element conventional meta-analyses lack: a clear causal interpretation within a well-defined target population (see Section  \ref{sec:lack}). This emerging class of approaches is now referred to as \textit{causally interpretable meta-analysis}.


This article can be viewed as part of a broader movement to reinterpret meta-analysis through a causal lens. However, we deliberately take a step back from recent generalization frameworks that assume access to  external covariate data. Instead, we remain within the more traditional scope of meta-analysis, where only limited aggregate data (typically contingency tables of treatment by outcome) is available. Our goal is to estimate the causal effect of a treatment on a target population that can be expressed \revised{as a convex combination of all trial populations}.

We demonstrate that, when treatment effects are measured using the risk difference
, classical meta-analysis estimators admit a causal interpretation. However, this interpretation breaks down for nonlinear effect measures such as the risk ratio or odds ratio. For such cases, we propose new aggregation formulas for combining risk ratios or odds ratios across trials. These formulas build on the collapsibility property \citep{huitfeldt2019collapsibility, greenland1999confounding}: for collapsible measures, the marginal effect on the target population can be recovered as a weighted average of stratum-specific effects, using weights \revised{that differ from inverse variance weights}. For non-collapsible measures like the odds ratio, we show that it is still possible to recover a meaningful marginal effect for the target population \revised{using an arm-based approach}.

\revised{
Arm-based approaches in network meta-analysis \citep{hong2016bayesian} and our framework shares similarity as both model outcomes at the level of study arms. This relates to the ongoing arm-based vs contrast-based debate \citep{dias2015absolute, white2019comparison}, where contrast-based methods is often favored for respecting the randomization, but sometimes criticized for obscure interpretation under non-collapsibility. Classical arm-based models are primarily statistical; our approach goes one step further further by embedding this representation within causal inference, making the target population explicit. This goes in favor to arm-based formulations, as they align naturally with a well-defined causal interpretation.}

Our analysis reveals that classical meta-analysis approaches can sometimes lead to different (and even reversed) conclusions: a treatment may appear beneficial under standard techniques, yet prove harmful when evaluated from a causal perspective. These discrepancies\revised{, observed both in simulations and in experiments on real-world data,} can carry serious implications for public health decision-making (see in Sections \ref{sec:mismatch} and \ref{sec:realdata}).

\section{The standard approach: fixed-effects and random-effects meta-analysis} \label{sec:general}

For ease of exposition, let us consider a set of $K \in \bbN$ randomized controlled trials measuring the same binary outcome $Y$ (e.g. death) and evaluating the same treatments. We assume that the results of the studies are available in a tabular form:
\begin{table}[H]
\centering
\begin{tabular}{c|c|c}
 & $Y=1$ & $Y=0$ \\
	\hline
 $A=1$ & $n_{k}(1,1)$ & $n_{k}{(1,0)}$ \\
 	\hline
 $A=0$ & $n_{k}{(0,1)}$ & $n_{k}{(0,0)}$
\end{tabular}
\caption{A typical table summarizing the finding of the RCT $k \in [K]$}
\label{tab:setting}
\end{table}
\noindent where $A \in \{0,1\}$ denotes the treatment assignment ($A=1$ if treated and $A=0$ otherwise), and $n_{k}(a,y)$ corresponds to the number of individuals in the study $k \in [K]$ with treatment status $A=a$ and observed outcome $Y=y$. Based on these observations, we can easily compute for each study $k$ a contrast $\hat \theta_k$ and an estimated variance $\hat\sigma_k^2$. For instance, for the log-risk ratio, letting $n_k(a) = n_{k}(a,1)+n_{k}(a,0)$ be the total number of individuals with treatment status $A=a$, these quantities are given by
$$
\hat\theta_k = \log\frac{n_{k}(1,1)/n_k(1)}{n_{k}(0,1)/n_k(0)} \quad \text{and}\quad \hat\sigma_k^2 = \frac{1}{n_{k}(1,1)}-\frac{1}{n_k(1)}+\frac{1}{n_{k}(0,1)}-\frac{1}{n_k(0)},
$$
see for instance \citep[Section 5.2]{fleiss2013statistical}. Other common measures can be considered (risk difference, risk ratio, odds ratio, log-odds ratio, number needed to treat, etc), although in practice, the risk difference, log-risk ratio, and log-odds ratios are often used due to their convenient asymptotic properties, with simple asymptotic variances. 

The goal of meta-analysis is to aggregate the $K$ estimates $\{\hat\theta_k\}_{k=1}^K$ into a single global measure $\hat \theta$.
Traditional meta-analysis techniques usually rely on Gaussian models to aggregate the study findings \citep{tanriver2021comparison}. For instance, \textit{fixed-effects models} posit that
$
\hat\theta_k \sim \cN(\theta^*,\sigma_k^2),
$
for some unknown true effect $\theta^*$, with $\sigma_k^2$ well approximated by $\hat\sigma_k^2$. 
When one suspects heterogeneity across studies, 
one can resort to \emph{random-effects models}, which are hierarchical Gaussian models of the form
$$
\hat\theta_k|\theta_k \sim \cN(\theta_k,\sigma_k^2) \quad \text{and} \quad \theta_k \sim \cN(\theta^*,\tau^2).
$$
We will denote by $\theta^{\mathrm{RE}}$ (resp. $\theta^{\mathrm{FE}}$) the target estimand $\theta^*$ in the random-effect model (resp. in the fixed-effects model). The usual estimators for these estimands are given by a weighted sum of the $\hat\theta_k$ where the weights are proportional to the inverse of the (estimated) variance:
\begin{align*}
\hat \theta^{\mathrm{FE}} &= \sum_{k=1}^K \hat\omega_k \hat\theta_k \quad \text{where} \quad \hat\omega_k \propto \frac{1}{\hat\sigma_k^2} \quad \text{and}\quad  \sum_{k=1}^K \hat\omega_k = 1,\\
\hat \theta^{\mathrm{RE}} &= \sum_{k=1}^K \hat\omega_k \hat\theta_k \quad \text{where} \quad \hat\omega_k \propto \frac{1}{\hat\sigma_k^2 + \hat \tau^2} \quad \text{and}\quad  \sum_{k=1}^K \hat\omega_k = 1.
\end{align*}
In random-effects models, the between-study variance $\hat \tau^2$ is typically estimated using the DerSimonian-Laird estimator \citep{dersimonian1986meta}, restricted maximum likelihood methods \citep{viechtbauer2005bias}, or the Paule-Mandel estimator \citep{paule1982consensus}, among others. When we set $\hat \tau = 0$, we retrieve the fixed-effect model estimator.

\section{A causal framework for meta-analysis}\label{sec:causal}

The estimands of the previous section are purely statistical objects; in particular, they are not tied to a well-defined target population and intervention. 
To tackle this ambiguity, we now adopt a causal perspective and define the estimands of interest as a population-level causal effect. 
This perspective shifts the focus from mere statistical aggregation to the estimation of causal effects under clearly defined assumptions, thereby enhancing the interpretability of meta-analytic findings.

\subsection{Causal assumptions}

\begin{figure}[t]
    \centering
    \includegraphics[scale=.8]{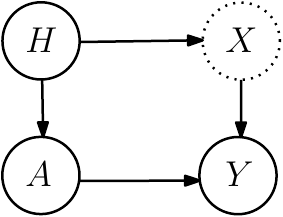}
    \caption{The directed acyclic graph (DAG) representing the causal structure of our model. The variable $X$ is not (or partially) observed.} 
    \label{fig:dag}
\end{figure}

We assume that an individual in each study can be described by a vector of covariates $X \in \cX$. We let $H \in [K]$ be the study membership, and $Y(a)$ be the counterfactual outcome of the individual under $A=a$.
\revised{We make three main assumptions:
\begin{as}[SUTVA] \label{as:sutva} $Y = Y(A)$.
\end{as}
\begin{as}[Collection of RCTs]\label{as:rct} $A \perp\!\!\perp Y(0),Y(1) \mid H$.
\end{as}
\begin{as}[No study-effect]\label{as:study}
The function $\mu_k$ defined as 
\begin{equation} \label{eq:exchangeability}
\mu_k(a,x) := \bbE[Y(a)\mid X=x,H=k] \quad \quad \forall a\in\{0,1\}, x\in\cX, 
\end{equation}
does not depend on the study $k$. We will simply denote it by  $\mu(a,x)$.
\end{as}
}The first two assumptions are standard in the causal inference litterature. Assumption~\ref{as:sutva}, (SUTVA, for \emph{Stable Unit Treatment Value Assumption}) states that each participant's outcome depends only on its own treatment assignment and that the treatment is the same for all participants. The second assumption (collecion of RCTs) is implied whenever the treatment has been randomly assigned within each study. 
\revised{The last Assumption~\ref{as:study} (\emph{no study-effect}) states that similar individuals in different studies respond similarly to the control or the treatment. Consequently, any observed heterogeneity between studies arises solely from differences in the covariate distributions. We will sometimes refer to the function $\mu(0,\cdot)$ as the \emph{baseline} function (the response to the control of an individual in any study), and $\mu(1,\cdot)$ as the \emph{response} function.} In the generalization literature, Assumption~\ref{as:study} is often coined \emph{exchangeability in mean}, \emph{no center-effect} \citep{robertson2021center, khellaf2025federatedO}, \revised{or sometimes \emph{weak response consistency} \citep{sobel2017causal}}. The causal graph implied by these assumptions is described in Figure~\ref{fig:dag}.

\revised{
\begin{rem}[About the \emph{no study-effect} assumption] The \emph{no study-effect} assumption is a strong structural assumption. Because \cite{sobel2017causal} work with individual participant data, they are able to relax this assumption and decompose heterogeneity more precisely into response inconsistency, treatment nonequivalence, selection effects, and related mechanisms. In our meta-analysis framework, however, the use of aggregated data prevents such precise modeling of heterogeneity, making stronger structural assumptions on the causal model unavoidable. It is important to note that the \emph{no study-effect} assumption can be reasonable in practice: in their case-study, \cite{sobel2017causal} find no evidence against weak response consistency.
\end{rem}
}



\paragraph{Observed data.} We assume that there is an underlying process of $n \in \bbN$ i.i.d. random variables of the form 
$\{(H_i,X_i,A_i,Y_i), i \in [n]\}$
whose common joint distribution satisfies Assumptions~\ref{as:sutva} to \ref{as:study}. Like in the usual meta-analysis setting, we assume to only have access to the aggregated statistics
$$
n_{k}(a,y) := \#\{i \in [n]~|~H_i = k, A_i = a, Y_i = y\}.
$$
\revised{We recall that $n_k(a) := n_{k}(a,1) + n_{k}(a,0)$ is the total number of participants having received treatment $a$ in the study $k$, and that $n_k = n_k(0)+n_k(1)$ is the total number of participants in study $k$.}



\subsection{Causal estimands} \label{sec:est}

In what follows, we only consider first-order \emph{population-level contrasts} 
(as defined in \cite{fay2024causal}), meaning that $\theta_k$ only depends on $\bbE[Y(0)\mid H=k]$ and $\bbE[Y(1)\mid H=k]$. 
In all generality, we will write
\begin{equation} \label{eq:defthetak}
\theta_k = \Phi(\bbE[Y(1)\mid H=k],\bbE[Y(0)\mid H=k]),
\end{equation}
for some function $\Phi$. These measures include the usual \revised{\emph{causal}} risk difference $\theta_k = \bbE[Y(1)\mid H=k]- \bbE[Y(0)\mid H=k]$ for $\Phi(x,y)=x-y$, the \emph{causal} risk ratio $\theta_k = \bbE[Y(1)\mid H=k]/\bbE[Y(0)\mid H=k]$ for $\Phi(x,y)=x/y$, its log version $\theta_k = \log(\bbE[Y(1)\mid H=k]/\bbE[Y(0)\mid H=k])$ for $\Phi(x,y) = \log(x/y)$, the \emph{causal} odds ratio for $\Phi(x,y) = x/(1-x)\cdot (1-y)/y$, etc. 
In the remainder of this paper, when clear from context, we may drop the adjective \emph{causal} when referring to these quantities and simply refer to them as risk difference, risk ratio, etc.

One can always express these quantities in term of baseline and response functions as defined in the previous section. Under the no-study effect assumptions, letting $\mu=\mu_k$ as in Eq~\eqref{eq:exchangeability}, and \revised{using the law of total expectation, one finds that
\begin{align*}
\bbE[Y(a)\mid H=k] &= \bbE\big[\bbE[Y(a)\mid H=k,X]\mid H=k\big] =  \mathbb{E}[\mu(a,X)\mid H=k] \\
&= \int \mu(a,x) \diff P_k(x),
\end{align*}
where population $P_k = \cL(X|H=k)$ is the distribution of the covariates in study $k$.} Therefore, $\theta_k$ as defined in Eq \eqref{eq:defthetak} depends on 1) the function $\mu$, and 2)  the study $k$ through the distribution $P_k$. We will denote this function by $\theta(\cdot,\cdot)$, so that $\theta_k = \theta(\mu,P_k)$, and we may drop $\mu$ from the notations when clear from context.  We now come to the central definition of this paper, stated here in an informal manner to ease the exposition, and in a formal way in Appendix~\ref{app:def} (Definition~\ref{def:interp}). 
\revised{
\begin{definition}[Informal] \label{def:interp_info}
We say that an estimand $\theta^*$ is \emph{causally interpretable} if there exists a target population $P^*$ such that $\theta^* = \theta(\mu,P^*)$ for all functions $\mu$.
\end{definition}

We impose that $\theta^* = \theta(\mu,P^*)$ must hold true for all outcome function $\mu$.  This implies in particular the target population $P^*$ is invariant under outcome pertubation (altering the values of the $Y$'s), treatment switch (replacing $A=1$ with $A=0$), etc. 

\begin{ex} To illustrate this definition, consider the case where
$$
\theta^* = \frac{1}{2}\theta(\mu,P_1) + \frac{1}{2}\theta(\mu,P_2),
$$
for two population distributions $P_1,P_2$. When $\theta$ is the risk difference, that is
$$
\theta(\mu,P) = \int \left(\mu(1,x)-\mu(0,x)\right) \diff P(x),
$$
then one can easily check that
$$
 \theta^* = \int \left(\mu(1,x)-\mu(0,x)\right) \diff P^*(x) = \theta(\mu,P^*) \quad\text{with}\quad P^* = \frac12 P_1 + \frac12 P_2. 
$$
This equality holds true for all functions $\mu$, so that $\theta^*$ is a causally-interpretable estimand per Definition~\ref{def:interp_info}. We refer to Section~\ref{sec:lack} for a throughout discussion of this property, including a counter-example of a non-causally interpretable estimand in the case of the log-risk ratio.
\end{ex}}

We argue that causal interpretability is a crucial property for any aggregate measure, particularly in policy-making contexts, where the primary objective is to estimate the effect of a treatment on a clearly defined population. Preserving this interpretability when combining general causal measures requires new estimators.

\subsection{Aggregation formulas for causal meta-analysis} \label{sec:aggr}

The recent  line of research commonly referred to as \emph{causally interpretable meta-analysis} \citep{dahabreh2023efficient} also seeks to cast meta-analysis in a causal framework.
However, these works focus on defining a target population $P^*$ in terms of an explicit external covariate distribution, and resort to reweighting or transport strategies of individual level patient data to estimate $\theta(P^*)$ \citep{rott2024causally}. In other words, they reframe meta-analysis as a \emph{generalization} problem of causal contrast \citep{colnet2024causal, boughdiri2025unified, dahabreh2024learning}.
Other related works aim at estimating the treatment effect within a specific RCT population \citep{robertson2021center} using multiple RCTs. 

\revised{
In contrast, in our meta-analysis framework, we have access only to the aggregate quantities reported in Table~\ref{tab:setting}, which is typically the information available from study reports. While this limited level of detail prevents the direct application of individual-level generalizations of causal contrasts, our approach remains applicable at the aggregate level. Note however that our analysis is restricted to target populations that can be represented at the same level of aggregation, i.e., as mixtures of the study populations:
\[
P^* := \sum_{k=1}^K \alpha_k^* P_k,
\]
for some prescribed weights $\alpha_k^*$ summing to $1$. Typical choices for $\alpha_k^*$ are:
\begin{enumerate}
    \item $\alpha^*_k = 1/K$: uniform weights for all studies;
    \item $\alpha^*_k = \bbP(H=k)$: corresponds to $P^* = \cL(X)$, that is, considering the pooled population of all studies;
    \item Setting $\alpha^*_k$ to balance covariate profile, in the context where one has access to the so-called \emph{Table~1} for each study, that is, statistics that summarize the covariate distribution $P_k$.
\end{enumerate}}
Depending on the context, one may need to estimate $\alpha^*_k$, and we let $\hat\alpha_k$ be the corresponding estimator. For instance, in the case of uniform weights, one can simply set $\hat\alpha_k = 1/K$. When $\alpha^*_k = \bbP(H=k)$, one may choose $\hat\alpha_k = n_k/n$. 

We let 
$$
\psi(a) := \int \mu(a,x) \diff P^*(x)
$$
be the mean outcome among the treated $(a=1)$ or among the control $(a=0)$ in the population $P^*$. 
Using the fact that $\int \mu(a,x) \diff P(x)$ is linear in $P$, we can estimate the quantity $\theta(\mu,P^*)=\Phi(\psi(1), \psi(0))$ through
\begin{equation} \label{eq:defcausal}
    \hat\theta^{\rm causal} := \Phi(\hat\psi(1),\hat\psi(0))\quad\text{with}\quad \hat\psi(a) := \sum_{k=1}^K \hat\alpha_k \hat\psi_k(a),
\end{equation}
and where $\hat\psi_k(a) = n_{k}(a,1)/n_k(a)$. This contrasts with classical fixed-effects (FE) and random-effects (RE) meta-analysis methods that aggregate estimates at the study level, i.e.
$$
\hat\theta^{\rm FE/RE} = \sum_{k=1}^K \hat\omega_k\Phi(\hat\psi_k(1),\hat\psi_k(0)) \quad \text{v.s.} \quad \hat\theta^{\rm causal} = \Phi\left(\sum_{k=1}^K\hat\alpha_k\hat\psi_k(1),\sum_{k=1}^K\hat\alpha_k\hat\psi_k(0)\right).
$$
In other words, instead of computing a weighted average of study-specific contrasts with weights inversely proportional to their estimated variances, causal meta-analyses aggregate each arm separately using the weights $\hat\alpha_k$, and then apply the function  $\Phi$ (recall that e.g. $\Phi(x,y) = x-y$ for the risk difference, $\Phi(x,y) = x/y$ for the risk ratio) to the aggregated arm-level estimates. 

\revised{
\begin{thm} \label{thm:consistent} Assume  that the weights $\hat\alpha_k$ are consistent estimators of $\alpha_k^*$ and that $\Phi$ is continuous at $(\psi(1),\psi(0))$. Then $\hat\theta^{\rm causal}$ is a consistent estimator of $\theta(P^*)$.
\end{thm}}
A strength of classical meta-analysis methods is that they readily come with an estimate of the variance of their estimator. The same is true for causal meta-analysis. We derive below the formulas for estimating the asymptotic variance of $\hat\theta^{\rm causal}$ for the \emph{pooling} weights $\hat\alpha_k = n_k/n$.

\revised{
\begin{thm} \label{thm:variance} Assume that $\alpha^*_k = \bbP(H=k)$ and $\hat\alpha_k = n_k/n$. Assume furthermore that $\Phi$ is differentiable in a neighborhood of $(\psi(1),\psi(0))$. Then $\sqrt{n}(\hat\theta^{\rm causal}- \theta(P^*))$ converges in law towards a centered Gaussian. Denoting by $\sigma^2$ the variance of this asymptotic law, a consistent estimator of $\sigma^2$ is given by 
$$
\hat\sigma^2 := \hat\sigma^2(1)  \hat v(1)^2 +  \hat\sigma^2(0) \hat v(0)^2 + 2 \hat\gamma \hat v(1) \hat v(0),
$$
where $(\hat v(1),\hat v(0)) := \nabla \Phi(\hat\psi(1),\hat\psi(0))$ and where
\begin{align*}
    \hat\sigma^2(a) &:=  \sum_{k=1}^K \frac{n_k^2}{n n_k(a)} \hat\psi_k(a)(1- \hat\psi_k(a)) +\sum_{k=1}^K \frac{n_k}{n}\hat\psi_k(a)^2-\hat\psi^2(a), \\
  \text{and}   \qquad\qquad \hat\gamma &:=  \sum_{k=1}^K \frac{n_k}{n} \hat\psi_k(1) \hat\psi_k(0) - \hat\psi(1) \hat\psi(0).
\end{align*}
\end{thm}}

The proofs of these two theorems are given in Appendix \ref{app:proofs}.

\paragraph{\revised{Link with collapsibility.}} When the target population is $P^* = \cL(X)$, one way to identify the target contrast $\theta^{\mathrm{causal}}$ is to use \emph{collapsibility} formulas \citep{colnet2023risk} by stratifying on the variable $H$. In the case of the risk difference or the risk ratio, these formulas take the following form: 
$$
\theta_{\mathrm{RD}}(\mu,P^*) = \bbE\left[\theta_{\mathrm{RD}}(\mu,P_H)\right] \quad\text{and}\quad\theta_{\mathrm{RR}}(\mu,P^*) = \bbE\left[\frac{\bbE[Y(0)|H]}{\bbE[Y(0)]} \theta_{\mathrm{RR}}(\mu,P_H)\right],
$$
where the outer expectation is taken with respect to $H$. \revised{For a precise definition of collapsibility, as well as the derivation of these formulas, we refer to \cite[Sec.~3.3]{colnet2023risk}.} These identification formulas, in turn, suggest new estimators:
\begin{align*}
    \hat \theta^{\mathrm{causal}}_{\mathrm{RD}} &= \sum_{k=1}^K \frac{n_k}{n} \hat \theta_{\mathrm{RD},k} \quad \text{where} \quad  \hat \theta_{\mathrm{RD},k} = \hat \psi_k(1) -\hat\psi_k(0), \\
    \text{and} 
    \quad \theta^{\mathrm{causal}}_{\mathrm{RR}} &= \sum_{k=1}^K \frac{n_k}{n} \frac{\hat\psi_k(0)}{\hat\psi(0)}\hat \theta_{\mathrm{RR},k} \quad \text{where} \quad  \hat \theta_{\mathrm{RR},k} = \frac{\hat \psi_k(1)} {\hat\psi_k(0)}.
\end{align*}
One can check that these formulas actually coincide with the ones of Eq~\eqref{eq:defcausal} when $\hat\alpha_k=n_k/n$ and with $\Phi(x,y)=x-y$ or $\Phi(x,y)=x/y$. Note that this formulation  yields a weighted mean of the per-study effects, similar to \emph{classical} meta-analysis, but with different weights. For the risk difference, the weights are given by the study proportions, whereas for the risk ratio they correspond to collapsibility weights, which depend on the baseline risk. Such a formulation is not possible for the odds ratio, since it is non-collapsible, although it remains estimable through an arm-based aggregation strategy, as shown in Eq \eqref{eq:defcausal} with $\Phi(x,y) = \frac{x}{1-x} \cdot \frac{1-y}{y}$.

\section{Comparing causal and classical meta-analysis} \label{sec:comparing}

\subsection{The lack of causal interpretability of classical meta-analysis} \label{sec:lack}

Analysing the causal interpretability of classical meta-analyses requires careful consideration, as the setting in which \emph{interpretability} is defined (i.e., Definition~\ref{def:interp_info}) differs from the statistical modeling of usual meta-analysis methods. In particular, in the random-effects model, the true per-study effects $\theta_k$'s are modeled as i.i.d. Gaussians, which is not the setting of the previous section. There are  two ways of bridging the gap between these two models.

\paragraph{\revised{Hierarchical models.}} In classical meta-analysis settings, the $\theta_k$'s are modeled as random variables, while they are treated as deterministic in our causal setting. Since $\theta_k$ only depends on the $k$-th study through its population ($\theta_k = \theta(\mu,P_k)$ as defined in Eq \ref{eq:defthetak}), one way to endow $\theta_k$ with randomness is to assume that the populations $P_k$'s have been previously drawn from a common prior distribution $\Pi$: 
$$
P_1,\dots,P_K \sim \Pi \quad \text{i.i.d.} 
$$
Under this assumption, the per-study effects are i.i.d. random variables and the random-effects estimand is simply the mean of these random variables:
$$
\theta^{\rm RE} := \bbE_{P_k\sim \Pi}[\theta(\mu,P_k)].
$$
If $\theta(\mu,P)$ depends linearly on $P$, as it is the case for the risk difference, then the expectation and $\theta$ commute and one can write 
$$
\theta^{\rm RE} := \bbE_{P_k\sim \Pi}[\theta(\mu,P_k)] = \theta(\mu,P^*) \quad \text{with}\quad P^* := \bbE_{P_k\sim \Pi}[P_k],
$$
so that a random-effect model indeed targets a causal quantity which is the effect of the treatment of the average population $P^*$. However, when $\theta(\mu,P)$ is not linear in $P$, then one cannot put $\theta^{\rm RE}$ under the form $\theta(\mu,P^*)$ (with $P^*$ independent of $\mu$) in all generality and the random-effects model loses its causal interpretability. We illustrate this fact with a simple counter-example below.

\revised{
\begin{ex} Consider the case where $\cX = \{0,1\}$, and assume that $\Pi$ only takes two values $P_1$ and $P_2$ with equal probability. Consider for instance that the estimand of interest is the log-risk ratio. We set $p_{k} := P_k(X=1)$. Then
\begin{align*}
    \theta^{\rm RE}_{\rm log-RR} &= \frac{1}{2} \log\left(\frac{p_1 \mu(1,1)+(1-p_1)\mu(1,0)}{p_1 \mu(0,1)+(1-p_1)\mu(0,0)}\right) + \frac{1}{2} \log\left(\frac{p_2 \mu(1,1)+(1-p_2)\mu(1,0)}{p_2 \mu(0,1)+(1-p_2)\mu(0,0)}\right). 
\end{align*}
We argue that, in general, the above expression cannot be written in the form
$$
\theta_{\rm log-RR}(\mu,P^*) = \log\left(\frac{p^* \mu(1,1)+(1-p^*)\mu(1,0)}{p^* \mu(0,1)+(1-p^*)\mu(0,0)}\right),
$$
with $p^* = P^*(X=1)$ independent of $\mu$. For instance, solving $\theta_{\rm log-RR}(\mu,P^*) = \theta^{\rm RE}_{\rm log-RR}$ for $p_1=0.1$, $p_2=0.9$, $\mu(1,1)=\mu(1,0)=0.5$, $\mu(0,1)=0.9$ and $\mu(0,0)=0.1$ yields $p^* \approx 0.36$. Using the same $p_1$, $p_2$, and $\mu(1,\cdot)$ but with $\mu(0,1)=0.1$ and $\mu(0,0)=0.9$ yields the same estimand $\theta^{\rm RE}_{\rm log-RR}$, but now $p^* \approx 0.64$.

More generally, we show in Appendix \ref{app:lack} that $\theta^{\rm RE}_{\rm log-RR}$ is a causally-interpretable estimand if and only if $p_1=p_2$.
\end{ex}}

\paragraph{\revised{Large-sample behavior.}} Although the quantities $\theta_k$ are not treated as random in our causal framework, the estimators $\hat\theta^{\rm FE}$ and $\hat\theta^{\rm RE}$ can be computed from the same aggregated data $\{n_{k}(a,y)\mid  a,y \in \{0,1\}, k\in [K]\}$. Accordingly, we could define $\theta^{\rm RE/FE}$ not as parameters of an additional statistical model, but as the large-sample limits of these estimators, when they exist:
$$
\theta^{\rm RE/FE} := \lim_{n \to \infty} \hat\theta^{\rm RE/FE}.
$$
When the weights $\hat\omega_k$ converge to scalars $\omega^* _k$, these limits take the form
\begin{equation}
    \theta^{\rm RE/FE} = \sum_{k=1}^K \omega^*_k \theta(P_k). \label{eq:lsl}
\end{equation}

As in the previous discussion, these quantities cannot generally be expressed as $\theta(P^*)$ for a univocally defined $P^*$. The counterexample presented earlier applies equally in this context.

\subsection{When causal and classical meta-analysis coincide}
\revised{
There are two limiting regimes where one can interpret the output of a classical meta-analysis in a causal fashion. }

\paragraph{\revised{No heterogeneity.}} 
If all the studies measure the same treatment effect $\theta_k = \theta^*$ for all  $k \in [K]$, then the estimand of classical meta-analysis has the clear interpretation of being the average treatment effect in any of the studies $k \in [K]$ (or any convex combination of the studies), and one can write $\theta^{\rm RE} = \theta^{\rm FE} = \theta(P_k)$ for any $k\in[K]$ and for any of the interpretations above.  In this case, the three estimators $\hat\theta^{\rm FE}$, $\hat\theta^{\rm RE}$ and $\hat\theta^{\rm causal}$ are distinct yet consistent estimators of the same quantity (for any choice of weighting scheme $\hat\alpha_k$). Among them, $\hat\theta^{\rm FE}$ can be shown to achieve the smallest variance \citep[see e.g.][Proposition~1]{khellaf2025federatedO}.

 \paragraph{\revised{Linear contrasts.}} In the large-sample interpretation of causality, if the weights $\hat\omega_k$ converge almost-surely towards $\omega_k^*$, and if the contrast $\theta$ is linear (as for the risk difference), then
 $$
 \theta^{\rm RE/FE} = \sum_{k=1}^K \omega^*_k \theta(P_k) = \theta(P^*) \quad\text{where} \quad P^* = \sum_{k=1}^K \omega^*_k P_k, 
 $$
 and the random-effects and fixed-effects meta-analyses admit a clear causal interpretation. 
 
 In the presence of true heterogeneity (that is, $\tau > 0$ in the random-effects model), and if the intra-study variances $\hat\sigma_k$ vanish, the random-effects weights become asymptotically proportional to $1/\tau^2$, that is $\omega_k^* = 1/K$ for all $k\in[K]$. This implies that, in this regime, a random-effects meta-analysis is equivalent to a causal meta-analysis with uniform weights $\alpha_k^* = 1/K$, corresponding to the average population across all studies.

Regarding the fixed-effect model, in the large-sample limit, it is expected that $\sigma_k^2 \propto 1/n_k$ so that $\hat \omega_k$ is proportional to $n_k/n$. Up to a normalization constant, this suggests that $\hat\omega_k \xrightarrow[]{\approx} \omega_k^* = \bbP(H=k)$. In other words, a fixed-effect meta-analysis can be interpreted as a causal meta-analysis with pooling weights $\alpha_k^* = \bbP(H=k)$. Interestingly, although fixed-effect models are derived under the assumption of no heterogeneity, this interpretation remains valid even when heterogeneity is present.

\subsection{A potential great mismatch} \label{sec:mismatch} 
We now compare classical and causal meta-analysis in a simulated setting, designed to illustrate the substantial discrepancies between the two approaches can arise in certain configurations. We implemented the three estimators $\hat\theta^{\rm FE}$, $\hat\theta^{\rm FE}$ and $\hat\theta^{\rm causal}$ on a synthetic dataset with two study populations $P_1$ and $P_2$. Results are shown in Figure~\ref{fig:mismatch}, and the data generation process is described in Appendix~\ref{app:dgp}.

\begin{figure}[H]
    \centering
    \includegraphics[width=\textwidth]{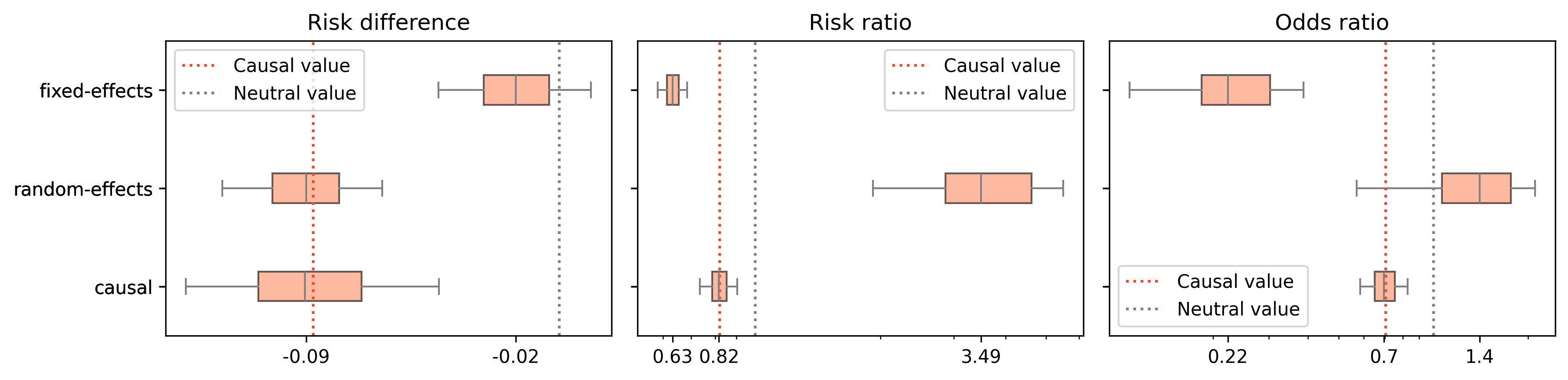}
    \caption{Comparison of a classical meta-analysis with fixed- or random-effects with our causally interpretable approach on a synthetic dataset, for three different causal contrasts (risk difference, risk ratio, odds ratio). The plots for the risk ratio and the odds ratio are given in log-scale. The grey lines correspond to the null-effect value associated with the given causal contrast ($0$ for the risk difference, $1$ for the risk ratio and odds ratio), and the red line corresponds to our causally interpretable estimand $\theta(P^*)$.}
    \label{fig:mismatch}
\end{figure}

For the causal method, we let the weights to be $\hat\alpha_k = n_k/n$, so that the target population corresponds to the two studies pooled together. Since both studies are balanced, this target population is simply $P^* = \frac12 P_1 + \frac12 P_2$.
As indicated by the red line in Figure~\ref{fig:mismatch}, assuming the outcome is desirable, the treatment has an overall harmful effect on $P^*$. For the risk difference, and as noted in the previous section, the random-effects estimator targets the same causal estimand $\theta(P^*)$ as the causal meta-analysis in this heterogeneous setting. In contrast, the fixed-effects meta-analysis clearly targets a different estimand, reflecting the limiting ratio of within-study variances $\hat\sigma_1 / \hat\sigma_2 \neq 1$.

Things change dramatically, however, when considering the risk ratio or the odds ratio.  On the same data, a random-effects meta-analysis misleadingly suggests that the treatment increases recovery by a striking three-fold, potentially leading to a recommendation in favor of the treatment, despite its harmful effect on the overall population. To understand why, we return to the expressions of the estimands $\theta^{\rm RE}_{\rm RR}$ and $\theta^{\rm causal}_{\rm RR}$. \revised{Because aggregation occurs on the log-risk ratio, $\theta^{\rm RE}_{\rm RR}$ can actually be written as 
$$
\theta^{\rm RE}_{\rm RR} = \exp\left(\theta^{\rm RE}_{\rm log-RR}\right) = \exp\left(\frac{1}{2}\theta_{\rm log-RR}(P_1)+\frac{1}{2}\theta_{\rm log-RR}(P_2)\right) = \sqrt{\theta_{\rm RR}(P_1)\theta_{\rm RR}(P_2)},
$$
where we used the large-sample interpretation of $\theta^{\rm RE}_{\rm log-RR}$ in the presence of heterogeneity (as defined in Eq \eqref{eq:lsl}). Letting $\psi_k(a) = \bbE[Y(a)\mid H=k]$, we have
\begin{align*}
\int \mu(a,x) \diff P^*(x) = \frac12 \int \mu(a,x) \diff P_1(x) + \frac12 \int \mu(a,x) \diff P_1(x) = \frac12 \psi_1(a)+ \frac12\psi_2(a),
\end{align*}
so that the estimands reduce to
$$
\theta^\mathrm{RE}_{\mathrm{RR}} = \sqrt{\frac{\psi_1(1)\psi_2(1)}{\psi_1(0)\psi_2(0)}} \quad\text{and}\quad \theta^\mathrm{causal}_{\mathrm{RR}} = \frac{\psi_1(1)+\psi_2(1)}{\psi_1(0)+\psi_2(0)}.
$$}
Crucially, there is no natural ordering between $\theta^\mathrm{RE}_{\mathrm{RR}}$ and $\theta^\mathrm{causal}_{\mathrm{RR}}$. In particular, $\theta^\mathrm{RE}_{\mathrm{RR}}$ is much more sensitive to small values of $\psi_0(k)$, where $\theta^\mathrm{causal}_{\mathrm{RR}}$ is more stable. This explains the dramatic difference observed in this synthetic example. 

\revised{
\begin{rem}
An important insight of our simulation is that the conclusions of a random-effects meta-analysis (i.e., whether the treatment is beneficial or harmful) can depend on the choice of causal contrast being aggregated. As illustrated in Figure~\ref{fig:mismatch}, the random-effects estimator yields a negative treatment effect for the risk difference, but a positive effect for the risk ratio. By contrast, our causal meta-analysis approach provides results that are provably consistent regardless of the causal contrast considered, highlighting its robustness.
\end{rem}}

\section{Causal meta-analysis on real-world data}  \label{sec:realdata}
\begin{figure}[H]
    \centering
    \includegraphics[width=\textwidth]{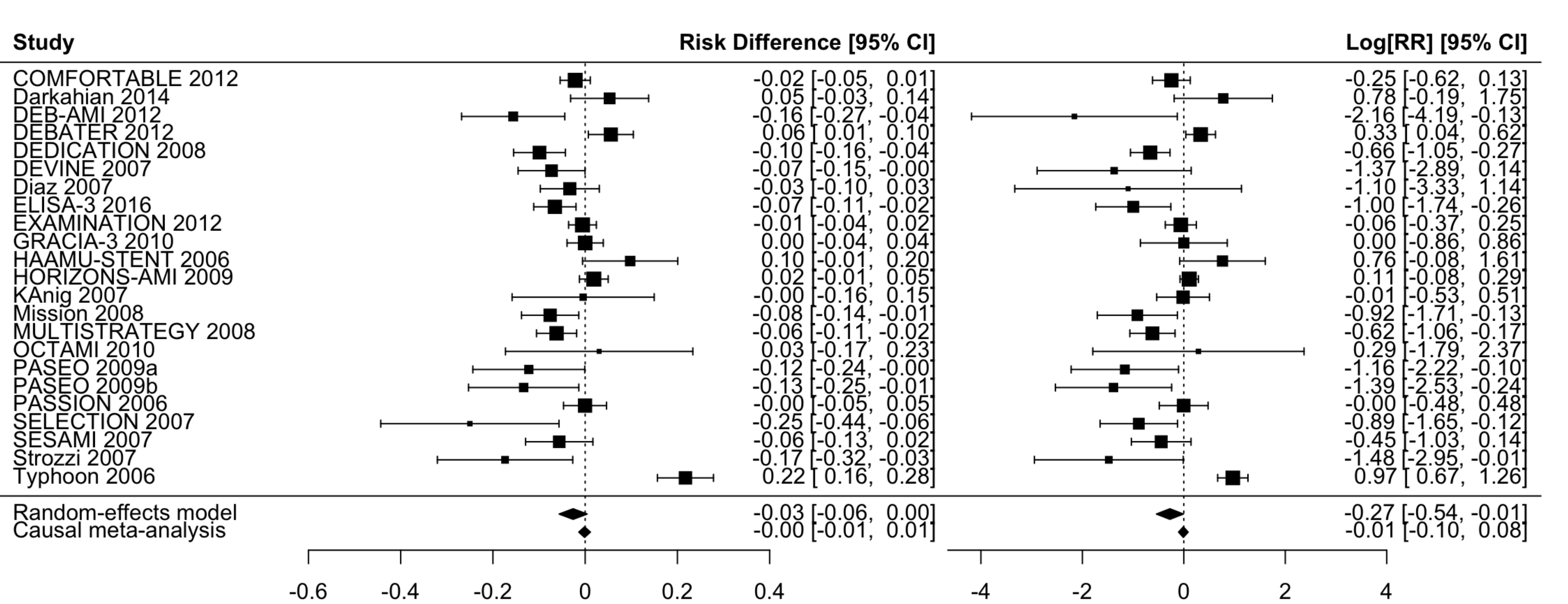}
    \caption{Comparison of a random-effects model with our causal approach on the data of Analysis 1.45 of \cite{feinberg2017drug}. The outcome is ``target vessel revascularisation''.}
    \label{fig:Meta2}
\end{figure}

We next illustrate how the differences between our causally interpretable meta-analysis and the classical random-effects approach manifest on real-world data.
We use data from the Cochrane Library \citep{cochrane_library}, focusing on a case study of drug-eluting versus bare-metal stents in acute coronary syndrome (Analysis 1.45 in \cite{feinberg2017drug}). As shown in Figure~\ref{fig:Meta2}, the random-effects model suggests a statistically significant treatment effect, whereas our causal method does not yield a definitive conclusion, underscoring how methodological choices can influence the findings. \revised{We stress that our goal is not to challenge the analysis of \cite{feinberg2017drug} but to use their dataset as an illustrative example. \cite{feinberg2017drug} already highlighted the limited amount of evidence available, and noted that the findings are highly sensitive to the method used for missing-data imputation.}

\begin{figure}[H]
    \centering
    \begin{subfigure}[b]{0.33\textwidth}
        \centering
        \includegraphics[width=\textwidth]{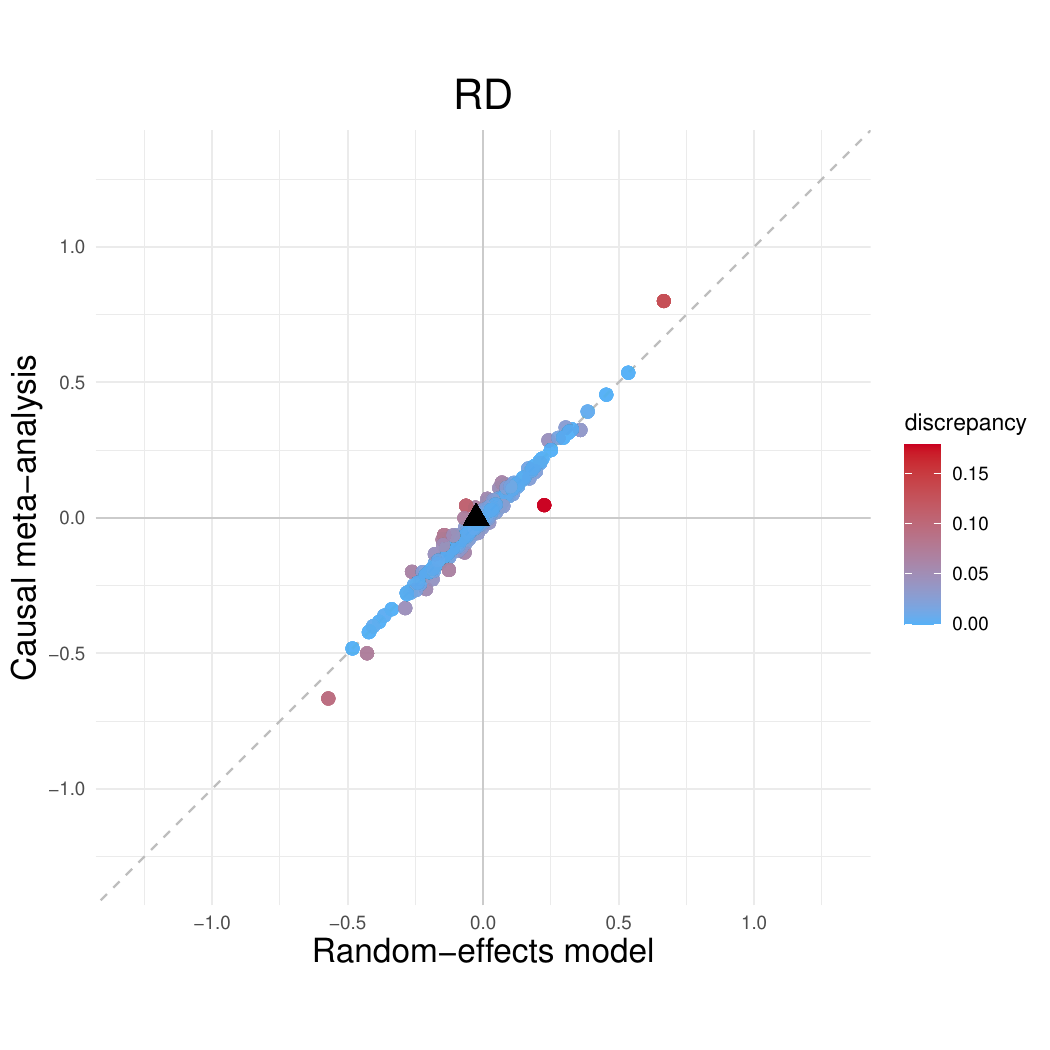}
        \label{fig:meta_RD}
    \end{subfigure}\hfill
    \begin{subfigure}[b]{0.33\textwidth}
        \centering
        \includegraphics[width=\textwidth]{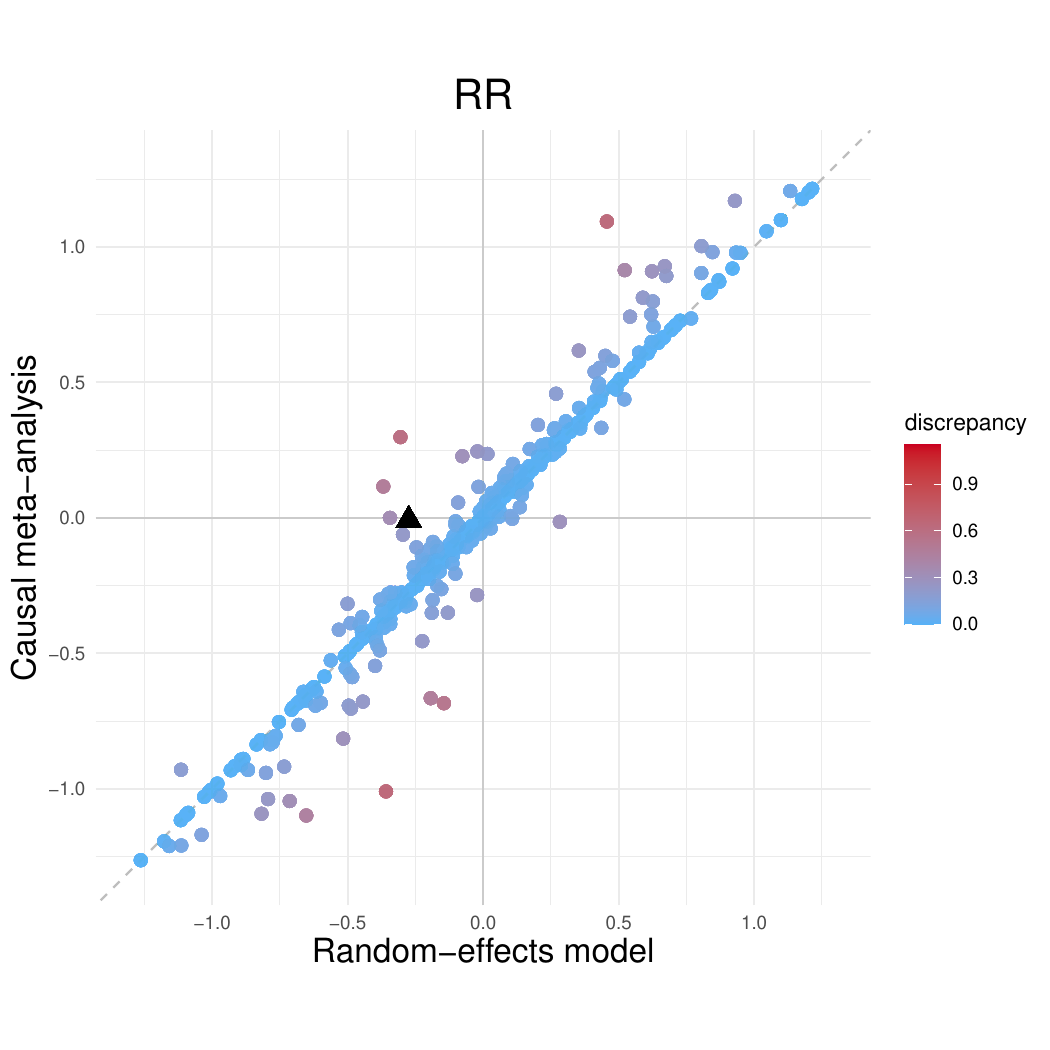}
        \label{fig:meta_RR}
    \end{subfigure}\hfill
    \begin{subfigure}[b]{0.33\textwidth}
        \centering
        \includegraphics[width=\textwidth]{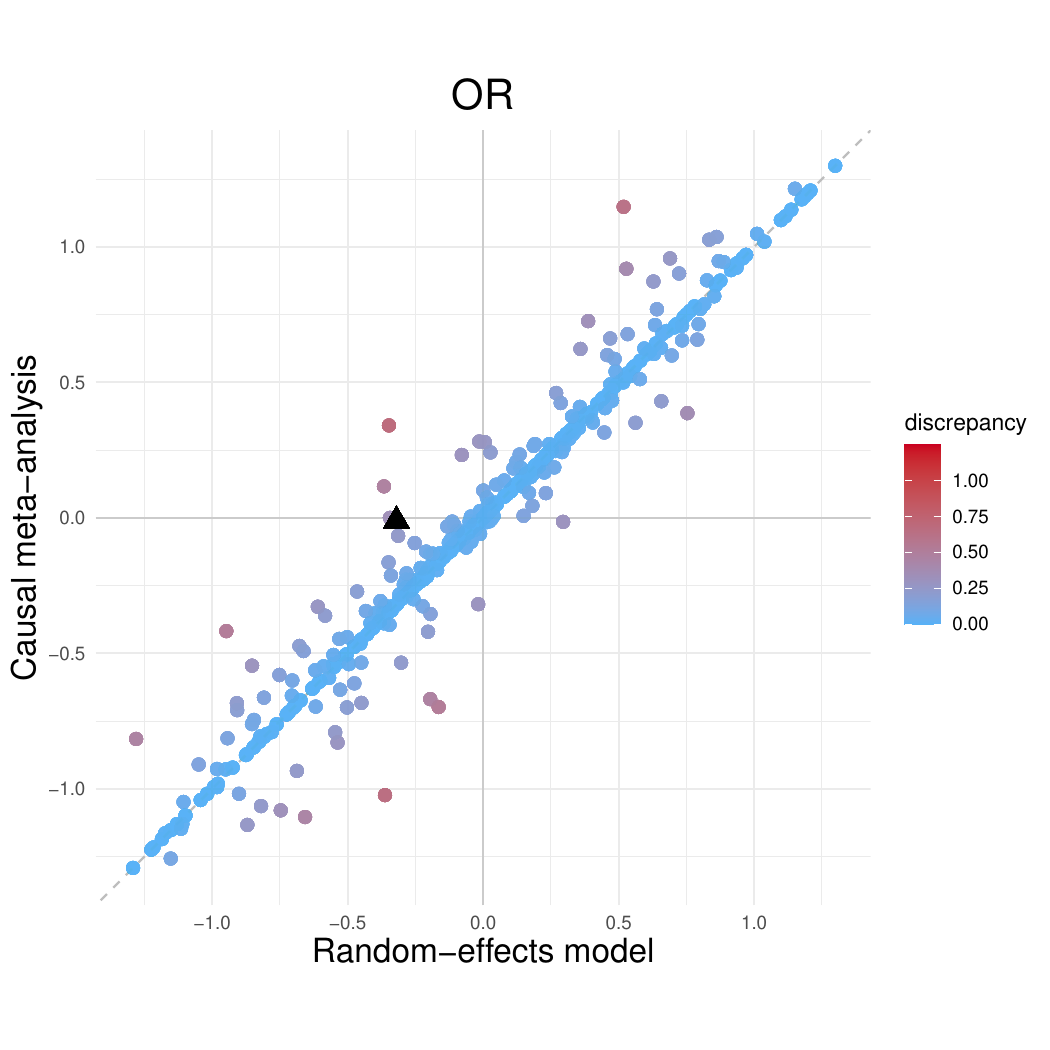}
        \label{fig:meta_OR}
    \end{subfigure}
    \caption{Comparison of a random-effects meta-analysis (x-axis) with the causal approach (y-axis) across 597 real-world meta-analyses from \cite{cochrane_library}. Point color represents the absolute difference between the two estimates. The meta-analysis highlighted in Figure~\ref{fig:Meta2} is marked with a black triangle.}
    \label{fig:meta_sidebyside}
\end{figure}

Extending beyond the single case study discussed above, we next examine how the two methods compare across a large collection of meta-analyses. Figure~\ref{fig:meta_sidebyside} presents three scatter plots comparing effect estimates across 597 meta-analyses from \cite{cochrane_library} for the three common causal contrasts: RD (risk difference), RR (risk ratio) and OR (odds ratio). Each point corresponds to a different meta-analysis, with the $x = y$ line indicating perfect agreement between the classical and causal approaches. The points generally cluster around this line, suggesting that the two methods often yield similar conclusions. Moreover, the clusters are roughly symmetric with respect to the $x=y$ line, indicating that neither method systematically produces higher or lower estimates than the other. Nonetheless, outliers like the example discussed in Figure~\ref{fig:Meta2}, highlighted by a black triangle in Figure~\ref{fig:meta_sidebyside}, show that discrepancies, although rare, can occur, particularly for non-linear contrasts. 

\revised{These differences are summarized in Table~\ref{tab:agreement_measures}. For each causal contrast, we report averages over the 597 meta-analyses for three quantities: 1) the absolute difference between the two methods' estimates, 2) the lengths of the confidence intervals produced by each method, and 3) the overlap of the two intervals, quantified by the Jaccard index (the ratio of the intersection length to the union length). Overall, the results show high average coverage, again reflecting strong agreement between the two approaches. Notably, the causal method tends to yield slightly narrower confidence intervals than the random-effects model.}

\revised{The random-effects analyses were conducted using the R-package {\tt metafor} \citep{metafor}, while the causal counterparts were computed using our R-package {\tt CaMeA} (Causal Meta-Analysis on Aggregated data), now available on CRAN \citep{camea}. 
}

\begin{table}[H]
\centering
{\small
\begin{tabular}{ccccc}
\toprule 
\small
{\bf Measure} & \small {\bf Discrepancy}  & {\bf \small CI length (RE)} & \bf \small CI length (causal) & \bf \small CI overlap (\%) \\
\hline
\bf RD & $0.01~(\pm 0.02)$ & $0.17~(\pm 0.17)$ & $0.14 ~(\pm 0.15)$ & $76.1\%~(\pm 23.2\%)$  \\
\hline
\bf RR & $0.07~(\pm 0.13)$ & $1.39~(\pm 1.28)$ & $1.22 ~(\pm 1.21)$ &  $80.9\%~(\pm 21.8\%)$  \\
\hline
\bf OR & $0.08~(\pm 0.15)$ & $1.74~(\pm 1.34)$ & $1.49 ~(\pm 1.27)$ & $80.4\%~(\pm 22.2\%)$  \\
\bottomrule
\end{tabular}}
\caption{Comparison between random-effects and causal meta-analysis. Reported values are averaged across the 597 meta-analyses, with standard deviations shown in parentheses.}
\label{tab:agreement_measures}
\end{table}



\section{Conclusion}

\revised{Adopting a causal perspective on meta-analysis provides a principled framework to clarify both the target population and the types of heterogeneity that can be addressed. Within this framework, the target population may represent the combined population across trials or a weighted combination of trial populations tailored to specific policy or health technology assessment objectives. 
By making assumptions and target populations explicit, this approach enhances the interpretability of meta-analysis findings for decision makers. The method is implemented in the R-package \texttt{CaMeA} (Causal Meta-Analysis on Aggregated Data), available on CRAN \citep{camea}.}

Our work deliberately focuses on a minimal-information setting, in which only summary-level data from randomized controlled trials are available. While analyses based on individual participant data can offer greater flexibility and efficiency \citep{morris2018meta}, such data are often inaccessible in practice due to regulatory, ethical, and logistical constraints, as well as the fragmented distribution of datasets across institutional silos.

Recent methodological advances have begun to address these challenges. Hybrid approaches that combine individual-level and aggregated data have been proposed to leverage partial access to richer datasets \citep{vo2025integration}. At the same time, federated causal inference frameworks \citep{khellaf2025federated, xiong2023federated, han2023multiply, pmlr-v235-guo24c, khellaf2025federatedO} have enabled decentralized estimation of treatment effects across distributed individual-level datasets. 
\revised{These developments point toward an emerging paradigm for evidence synthesis, in which information from multiple, heterogeneous sources, often subject to diverse access constraints, can be integrated within a coherent causal framework to produce robust and interpretable medical findings.}

 \subsection*{Disclosure Statement}
 The authors have no conflicts of interest to declare. Bénédicte Colnet is currently working for the Social Security (French Ministry of Health). This work represents the views of the author alone and was conducted independently during the author's personal time.

\subsection*{Acknowledgments} This work is part of the DIGPHAT project which was supported by a grant from the French government, managed by the National Research Agency (ANR), under the France 2030 program, with reference ANR-22-PESN-0017. It has also been done in the frame of the PEPR SN SMATCH project and has benefited from a governmental grant managed by the Agence Nationale de la Recherche under the France 2030 programme, reference ANR-22-PESN-0003.

We are deeply grateful to Raphaël Porcher, Rodolphe Thiébault and Tat-Thang Vo for their generous time and helpful remarks. We thank three anonymous referees for their constructive comments, which improved the manuscript.




\bibliographystyle{apalike}
\bibliography{references}

\appendix
\section{Appendix to Section \ref{sec:causal}} \label{app:causal}

\subsection{Definition of \emph{causally interpretable}} \label{app:def}

We denote by $Q$ the law of the counterfactual data $(H,X,A,Y(0),Y(1))$. We would like to say that an estimand $\theta^*_Q$ is causally interpretable if, as in the informal Definition~\ref{def:interp_info}, $\theta^*_Q$ can be written $\theta(\mu_Q,P^*)$ where $\mu_Q$ is the outcome function associated with $Q$, and where $P^*$ is target population. Since the space $\mathcal L(X)$ of distribution on $\cX$ can be large, equating $\theta_Q = \theta(\mu_Q,P)$ for a single distribution $Q$ may have infinitely many solutions, so that the interpretability of a solution $P^*$ might be vacuous. We thus tie the definition of being \emph{causally interpratable} to a statiscal model $\cQ$ on $Q$, that is a collection of probability distributions on the counterfactual data, and require that the equality $\theta^*_Q = \theta(\mu_Q, P^*)$ hold uniformly for $Q \in \cQ$. We allow $P^*$ to vary with $Q$ but we require it to stay invariant under perturbation the outcomes, as discussed in Section~\ref{sec:est}. We thus arrive to the following definition.

\begin{definition} \label{def:interp}
Let $Q$ be the law of $(H,X,A,Y(0),Y(1))$, and $\cQ$ be a statistical model on $Q$.  We denote by $Q_{\rm d}$ the marginal of $Q$ on $(H,X,A)$. We say that an estimand $\theta^*_Q$ is \emph{causally interpretable} relatively to a causal contrast $\theta(\cdot,\cdot)$ if there exists a map $P^* : Q_{\rm d} \mapsto P^*(Q_{\rm d}) \in \cL(X)$ such that $$\theta^*_Q = \theta(\mu_Q,P^*(Q_{\rm d})), \quad \forall Q \in \cQ$$
for all probability distributions $Q$.
\end{definition}
The subscript in $Q_{\rm d}$ stands for \emph{design} as it relates to the design of the meta-analysis and the designs and each study present in the meta-analysis. We give below a few examples and counter examples of causally interpretable estimand.
\begin{ex} \begin{enumerate}
    \item The estimand $\theta_Q^* := \theta(\mu_Q,P^*_0)$ for a given fixed target population $P^*_0$ is causally interpretable. 
    \item Letting $P_k := \cL_Q(X\mid H=k)$, and assuming that $H$ takes at least two values, then 
    $$
    \theta_Q^* := \frac12 \theta(\mu_Q,P_1)+\frac12 \theta(\mu_Q,P_2),
    $$
    is causally interpretable in the case of of the risk difference with $P^*(Q_{\mathrm d}) = \frac12 P_1 + \frac12 P_2$, as detailed in Section~\ref{sec:est}. For non-linear contrast however, it is not, in general, causally interpretable, as shown in Section~\ref{sec:lack}.
    \item One can also make the weight of the mixture of contrast depends on features of $Q_{\mathrm d}$. For instance, 
    $$
    \theta_Q^* := \sum_{k=1}^K Q(H=k) \theta(\mu_Q,P_k),
    $$
    is a causally interpretable for the risk difference with
    $$
    P^*(Q_{\rm d}) := \sum_{k=1}^K Q(H=k)  P_k = \cL_Q(X).
    $$
    One can also imagine more involved weights involving the distribution of $(X,A)$ conditional on $H=k$.
\end{enumerate}
\end{ex}

\subsection{Proofs of Section \ref{sec:aggr}} \label{app:proofs}
\begin{proof}[Proof of Theorem \ref{thm:consistent}] By the law of large number, it holds $n_{k}(a,1)/n \to \bbE[Y\ind\{H=k,A=a\}]$ and $n_k(a)/n \to \bbE[\ind\{H=k,A=a\}]$ almost-surely. The ratio $\hat\psi_k(a) = n_{k}(a,1)/n_k(a)$ thus converges to
$$
\frac{\bbE[Y\ind\{H=k,A=a\}]}{\bbE[\ind\{H=k,A=a\}]} = \bbE[Y(a)\mid H=k,A=a] = \bbE[Y(a)\mid H=k],
$$
where we used both Assumption~\ref{as:sutva} ($Y=Y(A)$) and Assumption~\ref{as:rct} ($Y(a) \perp\!\!\perp A \mid H$). Now assuming that $\hat\alpha_k \to \alpha_k^*$ almost-surely, we find that $\hat \psi_a$ goes to
$$
\sum_{k=1}^K \alpha_k^* ~\bbE[Y(a)|H=k] = \sum_{k=1}^K \alpha_k^* \int \mu(a,x) \diff P_k(x) = \int \mu(a,x) \diff P^*(x) = \psi(a),
$$
so that $\hat\theta^{\rm causal}$ goes to $\theta(P^*)$ almost-surely.
\end{proof}
\begin{proof}[Proof of Theorem \ref{thm:variance}] The vector  $(\hat\psi(1),\hat\psi(0))$ is straightforwardly asymptotically normal so that the first expression holds true for as long as $\hat\sigma^2(a)$ is a consistent estimator of the asymptotic variance of $\hat\psi(a)$ and $\hat\gamma$ is a consistent estimator of the asymptotic covariance between $\hat\psi(1)$ and $\hat\psi(0)$, per the $\delta$-method. We write
\begin{align*}
    \hat\psi(a) - \psi(a) &= \sum_{k=1}^K \frac{n_k}{n}(\hat\psi_k(a)-\psi_k(a)) + \sum_{k=1}^K \frac{n_k}{n}\psi_k(a) - \psi(a).
\end{align*}
Using that $\sqrt{n_k(a)} (\hat\psi_k(a)-\psi_k(a))$ goes to $\cN(0,\psi_k(a)(1-\psi_k(a))$ as $n \to \infty$, that $n_k /\sqrt{n n_k(a)}$ goes to $\alpha_k^*/\sqrt{e_k(a)}$ where $e_k(a) = \bbP(A=a|H=k)$, and that all terms in the sum are uncorrelated, we find that the first term, once re-scaled by $\sqrt{n}$ converges towards a Gaussian with variance
$$
\sum_{k=1}^K \frac{\alpha_k^2}{e_k(a)} \psi_k(a)(1-\psi_k(a)).
$$
The second term in the sum can be rewritten exactly $\frac{1}{n} \sum_{i=1}^n \psi_{H_i}(a) - \psi(a)$ so that its asymptotic variance, once rescaled by $\sqrt{n}$, is exactly 
$$\Var \psi_H(a) = \sum_{k=1}^K \alpha_k^* \psi_k(a)^2 - \psi(a)^2.$$
Finally, the two asymptotic variance can be added to each other are the two sums are uncorellated, as indeed, $\hat\psi_k(a)-\psi(a)$ is centered conditionally on $n_k$. The asymptotic variance of $\hat\psi(a)$ is thus given by
$$
\sum_{k=1}^K \frac{\alpha_k^2}{e_k(a)} \psi_k(a)(1-\psi_k(a)) + \sum_{k=1}^K \alpha_k^* \psi_k(a)^2 - \psi(a)^2,
$$
and the expression of the estimator $\hat\sigma(a)^2$ follows easily. Using the same decomposition as above, we find that 
$$
n \Cov(\hat\psi(1),\hat\psi(0)) = \Cov(\psi_H(1),\psi_H(0)) = \sum_{k=1}^K \alpha_k^* \psi_k(1) \psi_k(0) - \psi(1) \psi(0), 
$$
and the expression of $\hat\gamma$ follows. 
\end{proof}

\section{Appendix to Section \ref{sec:comparing}}

\subsection{Proof of Section \ref{sec:lack}} \label{app:lack}

We show that $\theta^{\rm RE}_{\rm log-RR}$ is a causally-interpretable estimand if and only $p_1=p_2$.

\begin{proof}
Assume that $\theta^{\rm RE}_{\rm log-RR}$ is a causally-interpretable estimand so that there exists $P^*$ such that $\theta^{\rm RE}_{\rm log-RR} = \theta_{\rm \log-RR}(\mu,P^*)$ for all $\mu$, and let $p^* = P^*(X=1)$. Setting
$\mu(0,1) = \mu(0,0) = \mu(1,1)= 1$ and $\mu(1,0)=0$, we find that $p_1p_2 = (p*)^2$. If now $\mu(0,1) = \mu(0,0) = \mu(1,0)= 1$ and $\mu(1,1)=0$, we get $(1-p_1)(1-p_2) = (1-p*)^2$. So that  $p_1 = p_2 = p*$.

The converse trivially holds.
\end{proof}

\subsection{Data generating process of Section \ref{sec:mismatch}}
\label{app:dgp}

The data consists of a iid realization of a tuples $(H,X,A,Y)$ where
\begin{enumerate}
    \item $H=1$ with probability $1/2$ and $H=2$ otherwise;
    \item $X | H = k \sim \cN(m_k,\eta^2)$. In the experiment, we set $m_1 = (1,0)$, $m_2 = (0,1)$ and $\eta = 0.1$;
    \item $A=1$ with probability $1/2$ regardless of the value of $H$;
    \item $Y|X,A=a$ is a Bernoulli of parameter $\sigma(X^\top \beta_a)$ where $\sigma(t) = e^t/(1+e^t)$. In the experiment, we set approximately $\beta_1 = (0.36,-1.38)$ and $\beta_0 = (2.94,-4.60)$. 
\end{enumerate}
For each meta-analysis, $n=1000$ patients are sampled. The experiment has been repeated 100 times.
\end{document}